# A Survey of Commodity WiFi Sensing in 10 Years: Current Status, Challenges, and Opportunities

Sheng Tan, Yili Ren, Jie Yang, Yingying Chen

*Abstract*—The prevalence of WiFi devices and ubiquitous coverage of WiFi networks provide us the opportunity to extend WiFi capabilities beyond communication, particularly in sensing the physical environment. In this paper, we survey the evolution of WiFi sensing systems utilizing commodity devices over the past decade. It groups WiFi sensing systems into three main categories: activity recognition (large-scale and small-scale), object sensing, and localization. We highlight the milestone work in each category and the underline techniques they adopted. Next, this work presents the challenges faced by existing WiFi sensing systems. Lastly, we comprehensively discuss the future trending of commodity WiFi sensing.

## I. INTRODUCTION

Recently, WiFi usage has been expanded from providing connection to desktops, laptops, and mobile devices to enabling Internet and network connectivity to smart and IoT devices. Such an expansion has resulted in the exponential growth of the available WiFi devices and the ubiquitous coverage of WiFi networks. This presents us the opportunity to widen the applications of WiFi from providing network communication to sensing the surrounding physical environment. When the WiFi signals or radio frequency (RF) signals travel through physical space, they interact with the objects or human body within the same environment. Those wave phenomenons include reflection, diffraction, and scattering, which cause the occurrences of multipath effects [1]. Such effects carry a rich set of information related to the surrounding physical environment such as the human motions and locations [2] as well as the status of the objects [3]. Indeed, there has been growing interest from research community to utilize multipath effects for various WiFi sensing applications, ranging from large-scale activity [2], [4], [5], [6] and small-scale motion recognition [7], [8], [9] to localization [10], [11], [12], [13], [14] and object sensing [3], [15], [16].

The WiFi-based sensing approach could be applied to a wide range of applications such as intrusion detection [17], [18], security and privacy [19], [20], [21], Human-Computer Interaction [22], [23], smart home [2], [15], mobile healthcare [24], [3], [25], and Augmented and Virtual Reality [26].

S. Tan is with the Department of Computer Science, Trinity University, San Antonio, TX, 78212.
Y. Ren and J. Yang are with the Department of Computer Science, Florida State University, Tallahassee, FL, 32306.
Y. Chen is with the Department of Electrical and Computer Engineering, Rutgers University, Piscataway, NJ, 08854.

In comparison to traditional sensing methods that rely on depth or infrared cameras and light sensors that are pre-installed within the environment [27], [28], [29], [30] and dedicated sensors such as RFID, gloves, motion sensors that are worn by the users [31], [32], [33], [34], [35], [36], [37], the WiFi sensing approach has many advantages. For instance, compared to wearable-based approaches, the users do not need to wear any dedicated sensors by using WiFi sensing systems. Moreover, compared to computer vision (CV) and acoustic-based systems [38], [39], [40], [41], [42], [43], [44], WiFi sensing approaches provide much better coverage and can work with non-line-of-sight (NLOS) scenarios because RF signals can sense through walls and pass through physical obstacles. Furthermore, commodity WiFi sensing can reuse existing WiFi devices at home without incurring an additional cost, and thus is promising for mass adoption for end-users in smart homes. Additionally, it reduces the potential privacy risk caused by computer vision-based systems.

Existing WiFi-based sensing systems either utilize commodity WiFi devices or specialized RF devices. The systems leveraging specialized devices, however, require to use of USRP software-defined radio and a specially designed receiver to extract carrier wave features that are not readily available on existing WiFi devices [45], [46]. Those systems are not suitable for large-scale deployment due to the requirement of specialized WiFi hardware and the high infrastructure cost. For this survey, we focus on the sensing systems utilizing commodity WiFi hardware. Such systems can enable ubiquitous sensing and large-scale deployments by reusing the WiFi infrastructures as well as leveraging the proliferation of WiFi devices and networks. Early WiFi-based systems [47], [18], [48], [49], [50], [51], [52] use received signal strength (RSS) to achieve coarse-grained sensing such as localization in simple environments. However, the popularity of Orthogonal Frequency Division Multiplexing (OFDM) technology-enabled WiFi devices makes it possible to extract more fine-grained channel responses known as Channel State Information (CSI), which reveals detailed measurements of each subcarrier compared to the RSS of the entire WiFi channel [53], [2], [54]. In this work, we mainly discuss the WiFi sensing systems leveraging CSI instead of RSS.

A number of surveys of general WiFi sensing have been published in recent years [55], [56], [57], [58], [59]. The main differences between this work and the above reviews are summarized as follows:
- A comprehensive review in the light of technical evolutions: most of the previous reviews focus on some specific sensing tasks without considering the technical evolutions from a higher perspective. This survey extensively reviews papers in the technical development path of WiFi sensing spanning over a decade.
- An in-depth analysis of current challenges: existing reviews usually focus on a very specific sensing task category. As different sensing tasks have totally different objectives and constraints, their challenges may vary from each other. Differently, this survey provides a thorough discussion of the challenges in general WiFi sensing.
- An extensive exploration of the potential future direction: the majority of the existing surveys lack a detailed examination of the future trends for WiFi sensing. This work provides a comprehensive discussion of various future directions utilizing WiFi sensing.

## II. BACKGROUND

### A. Channel State Information

In an indoor environment, the signal undergoes multipath propagation. Assuming there are $L$ different paths, the signal attenuation and delay on $l^{th}$ path is $a_l$ and $t_l$, respectively. The channel frequency response $h(f)$ can be described as following [60]:

$$h(f) = \sum_{l=0}^{L} a_l e^{-j2\pi f t_l},  \quad (1)$$

where $f$ represents the center-frequency. With 802.11n/ac systems, the WiFi NICs track fine-grained channel state information, which is a sampled version of the channel response including both phase and amplitude information. In particular, on the standard 20MHz WiFi channel, it measures the amplitude and phase for each of the 56 orthogonal frequency-division multiplexing (OFDM) subcarriers. With wider 40MHz channels, CSI measurements are available for 128 subcarriers. While received signal strength (RSS) measurements are a single quantity per packet that represents the signal-to-interference-plus-noise ratio (SINR) over the entire channel bandwidth, CSI contains amplitude and phase measurements separately for each OFDM subcarrier [2].

CSI is a metric that describes the channel properties of wireless communication links and considers the several factors affecting signal propagation, such as signal scatter, environmental attenuation, and distance attenuation. The purpose of the introduction of CSI is to ensure effective and reliable data transmission by quantifying the channel fading effect and adjusting the signal transmission rate. Specifically, when the wireless signal propagates in a multi-path manner, it will be obstructed by the objects in the line-of-sight (LOS) path, which leads to signal changes, including amplitude attention and phase shift. Besides, the reflection from the surrounding environment also changes the signal waveform. Therefore, CSI is introduced to evaluate the communication link state. That is to say, the quality of the wireless channel can be estimated by the CSI matrix, and the communication rate can be adjusted based on the CSI. In the IEEE 802.11n/ac standards, CSI is measured and parsed from the PHY layer using orthogonal frequency-division multiplexing technology. In the frequency domain, the wireless channel can be defined as:

$$Y = H \times X + N,  \quad (2)$$

where H is the channel matrix representing CSI information; the received and transmitted signal vectors are Y and X, respectively; N refers to an additive white Gaussian noise vector. Accordingly, $H$ can be expressed as:

$$H(i) = |H(i)|e^{jsin\angle H(i)},  \quad (3)$$

where $H(i)$ represents the value of CSI for the $i^{th}$ subcarrier which includes the amplitude and phase of the CSI; the amplitude and phase of the $i^{th}$ subcarrier are $|H(i)|$ and $\angle H(i)$, respectively.

### B. Signal Models

*1) Amplitude and phase:* The physical meaning of CSI amplitude is the quantification of the signal power attenuation after multi-path fading. Any motion in the WiFi-enabled area affects the wireless signal propagation and changes the amplitude of signal arriving at the receiver, leading to amplitude variation. A unique relationship can be derived between the change of amplitude and the motion magnitude as well as movement speed, which means the motions can be detected or even quantified using the measurement of amplitude. Many works utilize the amplitude for activity recognition [2], [22]. This shows amplitude has sensitivity over a wide scale of movements.

The phase measures the relative distance and direction of the signal propagation. Thus, it can be used to depict the signal changes and the corresponding motion [15]. However, as phase is periodic compared to amplitude and its measurement value is easily affected by device clock and carrier frequency. Therefore, it must be calibrated to remove noises for accurate motion and distance information extraction. Compared to simply using phase or amplitude, the combination of amplitude and phase can be leveraged and further improve the sensitivity and accuracy over activity recognition [4]. While the amplitude and the phase are sensitive to small movements in the physical environment, they cannot directly provide the spatial information (i.e., the spatial location of multiple movements or people in the 3D physical space).

*2) AoA and AoD:* The angle of arrival (AoA) could be leveraged to sense the movements together with the spatial information if the WiFi device is equipped with multiple antennas. Specifically, the angle of arrival (AoA) indicates the direction of the signal arriving at the receiver, and the angle of departure (AoD) indicates the direction of the signal departure from the transmitter. As we know, different propagation paths have different AoAs/AoDs, and when the signal from a propagation path is received across an antenna array, then the AoA/AoD will introduce a corresponding phase shift across the antennas in the array. Such a phase shift is a function of both the AoA/AoD and the distance between adjacent antennas. Therefore, both AoA and AoD have spatial resolution with respect to the target motion [26], [23].

In particular, the number of antennas at receiver/transmitter determines the resolution of the AoA/AoD estimates, respectively. Though it can be difficult to set up a large antenna array using commodity WiFi devices, existing work [61] combined the sampled channel response across all subcarriers (e.g., > 30) and multiple antennas (e.g., 3) to create a virtual sensory array. It is worth noticing, such an approach only works when the number of elements in the virtual sensory array is larger than the number of multipath components (i.e., signal reflections from surrounding objects/motions).

*3) ToF:* The propagation time the signal takes to travel along a particular path from the transmitter to the receiver is referred to as the time of flight (ToF). ToF estimation of a reflected signal defines an ellipse (with the transmitter and receiver as the two focal points) where the reflector is located. Similar to AoA/AoD, ToF also has spatial resolution or spatial information with respect to the target motion.

The resolution of ToF estimation is inversely proportional to the channel bandwidth. Therefore, the bandwidth is the major factor in determining time resolution and the distinguishable multipath components of the received signal based on the channel response. For example, given the widely used WiFi channel bandwidth at 5GHz is 40MHz, which yields a time of flight resolution of 25ns. Such a low bandwidth makes each received multipath component not resolvable due to the insufficient time resolution of each channel within a typical indoor environment. Existing work leveraged all the channels at 5GHz band (i.e., over 600MHz) by splicing those channels together [5] to increase bandwidth and further improve the ToF resolution. But such an approach requires scanning of over 20 available channels within the coherence time (i.e., less than 500ns) and solving convex optimization problems.

*4) Doppler Shift:* Movements of the transmitter, receiver, or reflectors all introduce frequency shifts to the carrier frequency of the signal, which is referred to as Doppler shift. Specifically, human motion causes a change in the length of the reflection path, resulting in frequency shifts. By measuring the signal frequency change, we are able to derive the direction, speed, and distance involved with human movement. Therefore, the Doppler shift has sensitivity over various motions [4], [5].

Doppler shift extraction is usually done by leveraging time-frequency analysis (e.g., Short-time Fourier transform (STFT), Discrete Wavelet Transform (DWT)). The resolution of Doppler resolution is a trade-off between frequency and time resolution. In the other words, the longer the interval, the finer the resolution but with lower time resolution. Thus, STFT has no guarantee of good frequency resolution and time resolution simultaneously. A long window length gives good frequency resolution but poor time resolution. The frequency components can be easily identified but the exact time when the frequency changes cannot be accurately determined. On the other hand, a short window length allows detecting when the signals change but cannot precisely identify the frequencies of the input signals.

*C. WiFi Sensing Datasets and Tools*

WiFi devices that support the IEEE 802.11n and OFDM could extract CSI values at the subcarrier level. Although CSI is included in WiFi since IEEE 802.11n, it is not reported by all off-the-shelf WiFi cards. The 802.11n CSI Tool [62] is the most widely used tool for CSI measurements extraction on commodity devices. It uses Intel 5300 WiFi cards to report compressed CSIs by 802.11n-compatible WiFi networks. It provides 802.11n CSI in a format that reports the channel matrices for 30 subcarrier groups, which is about one group for every 2 subcarriers at 20 MHz or one in 4 at 40 MHz. The Atheros CSI Tool [63] is another popular tool for CSI measurement extraction. It gives uncompressed CSIs using Qualcomm Atheros WiFi cards. For a 20MHz WiFi channel, the number of CSI subcarriers is 52 for the Atheros CSI Tool and 30 for the 802.11n CSI Tool. Recently, nexmon [64] has been developed to enable CSI extraction on a variety of WiFi devices including Nexus 5, Nexus 6P, Raspberry Pi B3+/B4, and Asus RT-AC86U. It supports 802.11a/(g)/n/ac with up to 80MHz bandwidth on the Broadcom WiFi chips.

It is crucial to build high-quality datasets that are available to the public that can enable continuous advancement of WiFi sensing-related research. In the past few years, there are several datasets have been released. For example, Yousefi *et al.* released a dataset that includes raw CSI measurements for 6 different users that perform the following activities: lie down, fall, walk, run, sit down and stand up [65]. A dataset of CSI samples for sign language recognition is provided by Ma*et al.* [66], which includes 276 and 150 sign words in various environments, such as home and lab. The other dataset is released with Widar and Widar2.0 which includes 80 CSI traces of a single user in the classroom, office, and corridor [67], [10]. Recently, another dataset released with Widar3.0 contains over 258K instances of CSI traces from 22 different gestures across 3 different environments [68].

## III. WIFI SENSING

In this section, we review the history of WiFi sensing in two aspects, including milestone work and the evolution of key techniques. In the past decade, the applications leveraging WiFi sensing can be generally divided into three main categories: activity recognition, object sensing, and localization. Moreover, activity recognition can be further categorized into large-scale activity recognition and small-scale gesture recognition. The road map for WiFi sensing systems in the past decade is shown in Fig 1.

*A. Milestones: Large-scale Activity Recognition*

In pervasive computing, it is an important task to provide accurate information on users' activities. Human activity recognition also known as HAR refer to the process of identifying the specific movement or action of a person. It plays a vital role in everyday life due to its ability in inferring high-level knowledge about human behaviors. By re-using existing WiFi devices, it is possible to sense human activities and further support a set of emerging applications (e.g., IoT, smart home, VR/AR), ranging from large scale movements including intrusion detection [18], daily activity recognition [2], and gait recognition [69] to small scale motions such as vital signs [24], and finger gesture [22].

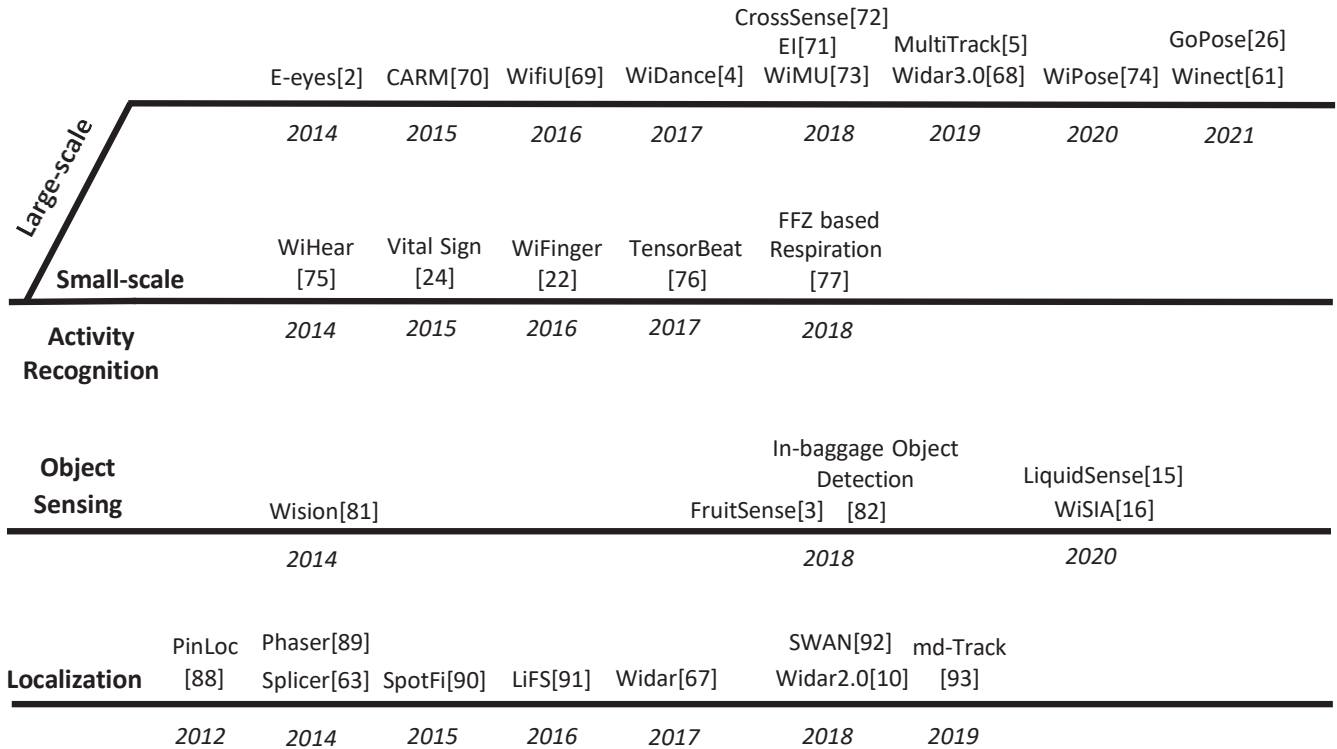

Fig. 1: A roadmap of WiFi sensing with milestone work.

Early system E-eyes [2] that proposed by Wang *et al.* in 2014 is the first work to recognize multiple activities at the same location in the home environment utilizing commodity WiFi devices. It performs device-free location-oriented activity identification through the use of existing WiFi access points and WiFi devices (e.g., desktops, thermostats, refrigerators, smart TVs, laptops). E-eyes leverages amplitude information to build location-activity profiles. It is capable of distinguishing various in-home activities where the profiles are affected by both the activities that people perform and the location people are located.

In 2015, Wang *et al.* proposed CARM [70], a CSI-based activity recognition system that aims to quantitatively correlate CSI dynamics and human activities using amplitude derived spectrogram. The proposed system attempts to quantify the correlation between CSI value dynamics and human movement speeds to further infer various activities. Similar work like WifiU [69] proposed in 2016 leverages the same approach to quantify the unique gait pattern of the individual user. But due to the lack of accurate phase information, the derived spectrograms from both systems can only quantify the intensity of the activity. Both systems can only extract the absolute value of Doppler shifts without arithmetic signs, thus, fail to identify the direction of motions. Later in 2017, WiDance [4] proposed by Qian *et al.*, is a ubiquitous gesture-based interaction interface using commodity WiFi. Such a system can extract accurate and comprehensive Doppler shifts with direction information from CSI that can accurately quantify different activities with both intensity and direction.

Different from early work mainly focusing on activity recognition for a specific setup, later work focuses on cross-domain problems including environment independent and multi-user compatible sensing. EI [71] proposed by Jiang *et al.* in 2018, is a deep-learning-based device-free activity recognition framework that can remove the environment and subject-specific information contained in the activity data. Specifically, it utilizes a Convolutional Neural Network (CNN) to extracts environment and subject independent features shared by the data collected on different subjects under different environments. Similar to EI, CrossSense [72] proposed by Zhang *et al.* in 2018 employs a machine learning algorithm to train, off-line, a roaming model that generates from one set of measurements synthetic training samples for each target environment.

Tan *et al.* proposed MultiTrack [5] utilizes Doppler shift based feature extraction to achieve environment independent activity recognition. It exploits the fact that the Doppler shift represents frequency change information of the movement, which wouldn't be affected by signal reflection from the static surrounding environment. By combining both deep learning neural network (DNN) along with velocity profile inferred from CSI that includes speed and direction information, Widar3.0 [68] proposed by Zheng *et al.* in 2019 is able to achieve cross-domain activity recognition (i.e., environments, locations and orientations of persons) without re-training. Differently,

Other concurrent works focus on providing multi-user support for the WiFi sensing system. In 2018, Venkatnarayan *et al.* proposed WiMU [73], a WiFi-based gesture recognition system that matches the generated virtual samples of desired gesture combination to the collected samples for multi-user compatible sensing. It, however, can only work when the system has pre-knowledge of all possible activities, which is infeasible in many real-world applications.

Later in 2019, MultiTrack [5] is capable of tracking multiple users and recognizing activities of multiple users performing them simultaneously. Different from WiMU, MultiTrack identifies and extracts the signal reflection corresponding to each individual user with the help of multiple WiFi links and all the available WiFi channels at 5GHz. Given the extracted signal reflection of each user, MultiTrack examines the path of the reflected signals at multiple links to simultaneously track multiple users. It further reconstructs the signal profile of each user as if only a single user has performed activity in the environment to facilitate multi-user activity recognition.

Later WiFi sensing systems aim to achieve free-form activity recognition that can extract more fine-grained information including the motion of each limb, joint, or hand/foot. Jiang *et al.* proposed a system that can construct a 3D human pose of many daily activities using commodity WiFi devices [74]. However, the number of poses that can be reconstructed is limited to only a set of predefined activities in the training phase. Recent system Winect [61] is a skeleton-based human pose tracking system for free-form activity estimation. Such a system does not rely on a set of predefined activities, thus can track free-form movements of multiple limbs simultaneously. Additionally, a system named GoPose [26] proposed by Ren *et al.* is a 3D skeleton-based human pose estimation system that offers on-the-go pose tracking in a home environment utilizing commodity WiFi.

### B. Milestones: Small-scale Activity Recognition

Different from large-scale activity recognition, small-scale gesture recognition mainly focus on applying existing WiFi sensing techniques to various applications. In 2015, Liu *et al.* [24] proposed a vital sign tracking system leveraging off-the-shelf WiFi during sleep. It is the first work to re-uses the existing WiFi network and exploit the fine-grained channel state information as opposed to coarse-grained RSS to capture the minute movements caused by breathing and heartbeats. Thus, it is possible to achieve widespread deployment and perform continuous long-term monitoring of both breath rate and heartbeats using the proposed system.

WiHear [75] proposed by Wang *et al.* is a system that enables WiFi signals to "hear" talks without deploying additional devices. It is done by detecting and analyzing fine-grained radio reflections from mouth movements based on pre-defined English vocabulary. By leveraging multiple antennas, WiHear explores the possibility of simultaneously tracking multiple people's talk.

In 2016, WiFinger [22] proposed by Tan *et al.*, is a system senses and identifies subtle movements of finger gestures by examining the unique patterns exhibited in the detailed CSI extracted from commodity WiFi devices. By devising the environmental noise removal mechanism, WiFinger mitigates the effect of signal dynamics due to the environmental changes. Moreover, by capturing the intrinsic gesture behavior, the proposed system can deal with individual diversity and gesture inconsistency.

TensorBeat [76] proposed by Wang *et al.* in 2017 is capable of estimating breathing rate for multiple persons. It is done by obtaining the CSI phase difference data between pairs of antennas at the receiver and by leveraging the tensor decomposition technique to achieve multi-user breathing rate estimation. Similar work proposed by Zhang *et al.* [77] utilizes the Fresnel diffraction model to quantify the relationship between the diffraction gain and the human target's subtle chest displacement to achieve respiration sensing. But those works can only do breath rate tracking without heart rate estimation. In 2020, Zeng *et al.* [78] proposed MultiSense that can achieve mutli-person respiration sensing while WiPhone [79] utilizing WiFi reflection from smartphone for respiration monitor.

### C. Milestones: Object Sensing

The majority of the existing WiFi sensing systems are focusing on human sensing applications [80], [7]. Nevertheless, many research efforts have been dedicated to object sensing. Specifically, object sensing refers to sensing the type, outline, or internal structure of a static object or a class of objects opposed to human motions. Early work like Wision [81] proposed by Huang *et al.*, demonstrates the possibility of WiFi imaging using 2.4GHz channels for daily objects such as leather couches and metallic shapes in LOS and NLOS scenarios. The proposed system, however, suffers from low image resolution and requires customized WiFi using a specialized device such as USRP. Furthermore, Wision can only work with objects made of a particular material that have good reflective properties (e.g., metallic surfaces), which greatly limits its application scenarios.

In 2018, FruitSense [3] proposed by Tan *et al.* demonstrates it is possible to sense the internal texture of an object (e.g., fruit) using commodity WiFi. It uses WiFi signals to enable non-destructive and low-cost detection of fruit ripeness. It leverages the larger bandwidth at 5GHz (i.e., over 600MHz) to extract the multipath-independent signal components to characterize the physiological compounds of the fruit and identify the fruit ripeness level (i.e., unripened, half ripen, ripen, over-ripen).

Other work proposed by Wang *et al.* [82] in 2018 leverage the fact that different material responses differently to WiFi signal to achieve detection of suspicious objects within the baggage that are suspected to be dangerous (i.e., defined as any metal and liquid object). It first detects the existence of the dangerous material type based on the reconstructed CSI complex value and then examines the object's dimension (i.e., liquid volume and metal object's shape) to determine the risk level of the object.

Recent work like LiquidSense [15] proposed by Ren *et al.* in 2020 can sense the miniature motions generated by high-frequency vibration and achieve liquid level sensing that can be applied to different daily liquids and containers in a smart home environment. The proposed system mounts a low-cost transducer on the surface of the container and emits a well-designed chirp signal to make the container resonant, which introduces subtle changes to the WiFi signals. By analyzing the subtle phase changes, LiquidSense can achieve liquid level detection. Moreover, WiSIA proposed by Li et al. [16] can achieve WiFi imaging by incorporating a cGAN (conditional Generative Adversarial Network) to enhance the boundary of different objects.

### D. Milestones: Localization

Localization is the process of determining the position of the subject of interest in space. Tracking seeks to identify the position of the subject of interest over time. In some cases, the problem of tracking may be reduced to a series of localization problems. Localization and tracking have been the topic of interest in many real-world applications [83], [84], [85], [86], [87], ranging from assisting technology (e.g., indoor and outdoor navigation) to facilitating HCI related systems such as virtual reality or augmented reality.

In 2012, Sen et al. proposed PinLoc [88], a fingerprint-based indoor localization system. This is the first work to explore the feasibility of leveraging channel state information extracted from the physical layer on commodity devices to achieve meter-level localization accuracy. A system like Phaser [89] proposed by Gjengset et al. in 2014 utilizes multiple antennas on the commodity WiFi devices to achieve phase calibration and improves device-based localization accuracy using an antenna array. But both systems require the hardware to be carried by the user in order for localization.

Splicer [63] proposed by Xie et al. in 2014 is a system that can derive higher resolution power delay profiles by splicing the CSI measurements from multiple WiFi frequency bands. By splicing up to 200MHz bandwidth allocated to 802.11n across 10 consecutive single band 20MHz channels, the ranging and localization error can be reduced to the sub-meter level.

In 2015, SpotFi [90] proposed by Kotaru et al. is another work that enables decimeter level localization using only commodity WiFi devices. Instead of relying on a large array of antennas, SpotFi combines the CSI values across subcarriers and antennas to jointly estimate the AoA and ToF of each path to further improve multipath resolution.

In 2016, Wang et al. proposed LiFS [91], a model-based device-free localization system that can also achieve decimeter level accuracy. Different from previous work, the proposed system leverages the unique characteristic of the power fading model and first Fresnel zone (FFZ) to model the localization problem. But it requires a specific setup that can be cumbersome and has limited applicable scenarios.

Widar [67] proposed by Kun et al. in 2017, is a system that can simultaneously estimate a human's moving velocity (i.e., speed and direction) and location. By utilizing Doppler frequency shifts to build a model that geometrically quantifies the relationships between CSI dynamics and human mobility, Widar can achieve decimeter-level localization accuracy. It is worth noticing, the model is based on the assumption human torso reflects more signals than other body parts and the prior knowledge of initial location information. A follow-up work Widar2.0 [10] has been proposed in 2018 to further extend the previous system's tracking and localization ability. Xie et al. proposed SWAN [92], a general antenna extension solution to commodity WiFi devices that can improve the performance of localization. Similar to Widar2.0, mD-Track [93] proposed by Xie et al. in 2019 continues to further push the resolution limitation of the current device-free WiFi localization and tracking system.

## IV. CHALLENGES

### A. Fine-grained Motion Sensing

Fine-grained motion sensing refers to the tracking of movements and their trajectories with a higher granularity that can be utilized to support more fine-grained HCI applications. Current commodity WiFi-based sensing systems mainly utilize the multiclass classification approach to achieve activity recognition. In general, multiclass classification is the process of classifying instances into one of multiple (i.e., more than three) classes. Such systems can only accurately recognize various activities that are well-defined and already included in the training samples. Thus, it is impossible to infer fine-grained motion trajectories only by utilizing the multiclass classification approach. Instead of using well-defined activity classes, recent works aim to tracking human motions [74], [61], [26] by reconstructing 3D human poses. But those systems still suffer from various shortcomings such as only work for predefined activities as well as limited sensing ranges. Currently, how to infer more fine-grained motion trajectory that can support both wider sensing range and better motion granularity is still an challenge.

### B. Multi-user Support

Multi-user support allows the WiFi sensing system to track multiple users simultaneously within the same environment (e.g., in the same room). By enabling multi-user compatibility, it is possible to apply such system to a wider range of applications under various scenarios. Existing commodity WiFi based sensing systems that is capable of multi-user tracking can only work well under specific application scenarios. For example, it is possible track multiple users' breath rates only if they are sleeping [8]. Other system aim to solve the problem by increasing the channel bandwidth [5] to achieve activity recognition and tracking. However, those approaches require high package transmission rate (i.e., over thousand pkt/s) which will disrupt normal network traffic. Furthermore, those systems can only separate very limited number of users (i.e., up to 3) and the tracking accuracy degrades quickly when the number of users keeps growing. Thus, it is still challenging to enable desirable multi-user support (e.g., a larger number of users, higher sensing resolution) on commodity WiFi devices.

### C. Complex Scene

The accuracy of the WiFi sensing systems has been improving over the last decade, but there are still many complex scenes where current systems can not perform well or have trouble dealing with. Firstly, most existing systems have a very limited sensing range and only consider smaller spaces (e.g., a single room in the office or home [2]). Such systems can not work well when apply to larger spaces with heavy pathloss and more complicated environments. For example, existing systems mainly focus on small indoor environments with one or two persons (e.g., living room, bedroom) and cannot work well when applied to a larger and more complicated public areas (e.g., large classroom, train station, bus station, and airport [5]). Additionally, existing work always assumes all the targets have similar size or scale but failed to consider the scenario where different targets might have various sizes or scales in the scene (e.g., adults, children and pets). Lastly, most systems only consider simpler sensing scenarios, which can not work well when the targets are overlapping or blocking others, resulting in target occlusion [61]. The aforementioned challenges greatly limit the applicable scenarios of the commodity WiFi sensing systems beyond sensing in smart home environments.

*D. Object Sensing*

Object sensing is part of the omnipresent component of modern sensing systems, which can be utilized by a wide range of applications. Different from activity recognition, localization and tracking, which mainly focus on the human subject and human motion, object sensing aims to infer the type, outline, or internal structure of a static object or a class of objects within the environment. The ultimate goal of object sensing is to provide the answer to the fundamental questions posted by various applications: what are the objects and has the object changed? Existing works leverage the WiFi imagining techniques to achieve object sensing can provide the outline information of the static object [81]. Another body of work attempts to sense the internal structure of the target objects [3], [15], [82] such as fruit ripeness level, liquid volume and suspicious object in the baggage. However, existing object sensing systems leveraging commodity WiFi can only provide limited resolution and are heavily customized for specific setup/tasks. Therefore, they can not be easily adopted by other applications. For example, current work can only sense a single object and cannot work well when dealing with a large number of objects with various sizes. With the advancement of sensing algorithms and commodity WiFi hardware, it will be interesting to see how to better solve the "general object sensing" problem.

*E. Deep Learning*

As a result of recent advancements in artificial intelligence research, especially deep learning algorithms, more and more WiFi sensing systems began to leverage various deep learning models (e.g., Artificial Neural Networks (ANN), Convolutional Neural Networks (CNN), RNN (Recurrent Neural Network)). By incorporating those deep learning models, existing systems can further improve accuracy and robustness while reducing processing latency [94], [95], [68], [71]. But there are still several limitations of embedding deep learning models into WiFi sensing systems.

A major limitation of the deep learning model is that it requires massive high-quality datasets for training. For example, due to the reliance on training samples, most deep learning-based sensing systems can not properly recognize other activities that do not include in the training data. Furthermore, the existing deep learning-based WiFi sensing systems lack labeled data for training. This is mainly because WiFi-based sensing data are difficult to label since those data are not intuitive and can not be easily labelled automatically. In addition, most deep learning models are usually customized to a particular domain or even a specific task. For example, because of the multipath characteristic of WiFi transmission, the received signals are heavily affected by the surrounding multipath environments. The data corresponding to the training signals could be very different from that of the testing due to the changes in the surrounding multipath environments. The sensing performance thus could suffer when the multipath environment changes (e.g., the system transferred to a new location, different user orientation, same activity performed by different users). Therefore, it is an open challenge on how to better incorporating deep learning models into current WiFi sensing systems.

V. FUTURE TREND

*A. Context Sensing*

In general, sensing tasks can be further divided into the following stages: detection, localization, recognition, and understanding. For detection, it aims to answer the question that if a particular subject present in the environment. This is usually the first stage of sensing because it only requires the system to detect the existence of a single subject or the occurrence of a single event. After the successful detection of the subject, the localization stage aims to find out its accurate location. Such a task not only requires the knowledge of the subject presented in the environment but also its specific location. Most of the commodity WiFi-based tracking and localization systems fall into this category. When detection and localization stages are done, the recognition stage attempts to detect and localize all the subjects present in the environment. Specifically, the system needs to achieve the localization of all subjects if they exist. For instance, the system that can achieve multi-user localization and tracking [5] or multi-user gesture recognition [73] utilizing commodity WiFi can be categorized into this level of the task.

Different from previous stages, the stage of understanding tries to recognize the subject in the context of the scene or surrounding environments, which is also known as context sensing. This is far more difficult to achieve compared to previous three stages due to the recognition results can vary depending on the different contexts. In other words, the same subjects might have a different meaning when such subject is under different contexts or in different environments. Take activity and gesture recognition as an example, the same human pose performed by various users or even the same user can be recognized as different activities or gestures under various contexts. Specifically, when the system senses a user is laying down on the bed, it is possible such user is sleeping, or he/she is listening to music or reading a book. It is extremely difficult for existing WiFi sensing systems to fully understand the intention without the context information.

In the real world, being aware of the context and surrounding environment is the key to achieving the goal of context sensing. The current commodity WiFi sensing system is still limited by the sensing capacity, which lacks the ability to truly understand the complex scene and utilize such information to further improve the sensing accuracy. With the continuous expansion of WiFi networks along with the growing number of WiFi-enabled devices, the opportunity to take advantage of ubiquitous WiFi and achieve context sensing can be further exploited in the near future.

*B. Semantic Sensing*

By including more important semantic properties and structures of scenes, current WiFi sensing systems is capable of supporting a wider range of applications. In particular, we define the term semantic sensing as the task to analyze a scene by considering the semantic context of its contents and the intrinsic relationships between them. In other words, the semantic sensing system is capable of understanding and describing the scene in a human-understandable form. Compare to context sensing, semantic sensing not only take the context information into consideration but can also infer more information of other possible contexts. Semantic sensing is remarkably challenging due to the complex interactions between objects, complex environments, different viewpoints and scale changes across different scenes, and the inherent ambiguity in the limited information provided by a given scene. It is still an open challenge even for well-developed computer vision or AI-based systems.

Considering the scenario of a family gathering with multiple family members spread through different rooms and performing various activities (e.g., playing, cooking, talking, watching TV). The existing WiFi sensing system is capable of detecting single-user motion within a specific room. For example, it is possible to infer if the user is watching TV in the living room [2]. Moreover, the current system can sense if kids are playing in the bedroom or two users are talking to each other in the dining room [5]. On the other hand, by combining the vital sign sensing [24] and activity sensing [61], future WiFi sensing system can infer the mood of the person who is watching TV. Moreover, through mouth motion and posture sensing, it is possible to perceive if the person in the conversation is agitating or not. In addition, by sensing both the activities of the children and their body language, we can detect potential bullying occurrences. Similarity, by leveraging object sensing [3], [15] and the motion of the user, the system can guide the person who is trying to cook a meal where the specific items are located in the kitchen, and if the restock of groceries or condiments is needed. Then, combining the multi-level sensing results of each person, human interactions and the surrounding environments, the system could describe the sense as a fun family gathering with lots of informative details. Thus, incorporating semantic sensing abilities into existing sensing system can be a long-standing goal and desirable research topics in the future.

*C. Privacy and Security*

The WiFi-based sensing systems can work with non-line-of-sight (NLOS) scenarios because RF signals can easily pass through physical obstacles such as walls. Due to the open nature of the RF signals, an adversary is capable of inferring a wide variety of information such as mouth motion, touch motion and vital signs without the awareness of the victim, especially by leveraging mobile devices. For example, WiHear [75] can be utilized to infer the highly confidential conversation in another room by sensing the mouth motion of the victim. WindTalker [96] allows the attacker to infer sensitive password information on mobile devices through typing motion. Moreover, the personal information such as vital sign [24], daily activity [2] and gait [69] can be eavesdropped by the attacker. Those information can be further utilized to achieve user identification or authentication [97]. By combing those information with other data such as location, it will cause serious consequences and raise privacy concerns.

Recently there are several work have been done that aim to solve the potential privacy issues mentioned above. For example, BPCloak [98] can erase the behavior information contained in the RF signals while preserving the ability of user authentication. IRShield [99] is capable of obfuscating the wireless channel to prevent attacker from eavesdropping and obtaining sensitive information of the victim. Meanwhile, it is possible to alleviate privacy concerns through passive and opportunistic WiFi sensing [6], [25]. Existing WiFi sensing systems that rely on active sensing model require user co-operation and customized traffic. Passive sensing model does not enforce user cooperation and take advantage of existing network environment without introducing additional traffic or active data collection.

WiFi sensing has also been used to further improve the security of wireless network. For example, Liu *et al.* [100] proposed a system that leverages channel state information for secret key extraction to ensure the confidentiality of wireless communication. Moreover, the fine-grained channel state information has been utilized to provide user authentication in the wireless network [19], [101]. In addition, systems have been proposed [86], [102] to locate the rogue access points in the WLANs. Since there is still limited number of research have been done in this direction, it will be interesting to see how future work address the existing privacy and security issue.

*D. Emerging Applications*

**Healthcare and Well-being.** With the advancements and increasing deployment of commodity WiFi networks and ubiquitous of WiFi-enabled devices, there is a grown interest in research in the healthcare domain utilizing WiFi sensing (e.g., vital sign and activity tracking for well-being, medicine dosage monitoring) [103]. With the integration of emerging technologies in the healthcare domain, WiFi sensing can help caretakers continuously monitor patients, record their wellbeing process, and report any acute situation in the case where abnormal behavior is detected. Thereby, it would be easier and more efficient to monitor and manage the lifestyles and well-being of patients with chronic diseases, the elderly people, the rehab taking patients, the patients dealing with obesity, the patients with cognitive disorders, and children. By utilizing object sensing, it is possible to closely monitor the dosage of medicine within the container and alter the user when it is time to take the medicine or avoid overtaking a certain medicine.

**Agriculture Applications.** Agricultural production is an important part of the global economy [104]. As the global population continues to grow, urbanization will lead to a continuous reduction in the area of arable land and the number of farmers. The agricultural production system faces many challenges. It is thus important to seek efficient and intelligent agricultural technologies, which can save manpower and material resources to promote high-quality and high-yield agricultural development.

Existing work has shown it is feasible to sense the physiological changes associated with fruit ripening for detecting the ripeness of various fruits utilizing commodity WiFi [3]. Thus it is possible to extend the WiFi sensing's capability from fruit sensing to crop monitoring. By utilizing WiFi sensing, it is possible to monitor the status of the crop in real time under various conditions (e.g., poor lighting or non-line-of-sight condition). Furthermore, WiFi sensing can also be applied to livestock monitoring which observes and analyze the behavior of livestock animals on the farm to better support farming needs.

**Other Applications.** The domain of logistics is an area of interest where WiFi sensing can be applied in the future [105]. This is because many logistics system requires real-time monitoring for better handling of various packages under different conditions. For example, it is possible to deploy a WiFi sensing system to track and monitor the condition of the items within the package both in storage or during the transportation process. It is also possible to utilize WiFi sensing to support smart city related applications [106]. Furthermore, WiFi sensing can be utilized to support a variety of emerging Human-Computer Interaction applications that demand the 3D human pose of free-form activity. For instance, virtual reality, medical training in extended reality, and existing smart home applications that require precision control as well as 3D free-form movement tracking can benefit from leveraging WiFi sensing.

*E. Sensing with Information Fusion*

It is possible to achieve fusion of the CSI data with other signal modalities, such as video, audio, Bluetooth, broadband cellular, ZigBee, GPS, motion sensor, and so on to achieve cross-domain sensing. For instance, it is possible to combine WiFi and camera together to improve the performance and robustness of the sensing system. This is because the CV-based approach can achieve higher performance with LOS as well as good lighting conditions, while the WiFi-based approach can work under NLOS and poor lighting conditions. Those sensing modalities can complement each other and boost the system performance under various conditions. Moreover, the camera data can be utilized for ground truth labeling during the training phase of a WiFi-based sensing system. Such an approach can greatly reduce the cost and human effort of ground truth labeling for WiFi sensing systems. Therefore, by investigating the potential of cross-modality data fusion, we are able to further improve the robustness and performance of the WiFi sensing system.

*F. Standard Dataset*

Almost all existing WiFi sensing systems conduct evaluation through specific data sample collection processes. The typical evaluation process includes the recruitment of multiple participants and asking them to perform a number of specified activities in the pre-set locations while the CSI data samples are collected during that period. Usually, the activities and experimental environment are determined by the nature of the applications and the system performance is only validated under its own arrangement. Therefore, it is very difficult to achieve comprehensive evaluation and comparison between different WiFi sensing systems due to the wide variety of experimental setups/conditions. Although there is work attempt to provide a comparison between several systems, the analysis and discussion are fairly simple. Such an issue greatly limited the advancement of WiFi sensing systems as a whole since each system mainly targets one specific setup. On the other hand, the standard datasets can be utilized to solve this problem. With the publicly available datasets, it is possible to assess the performance of the proposed systems from various aspects and achieve performance comparison between different systems. So far, there have not been many high-quality public datasets available. Thus, we consider the development of the open standard datasets to be one of the promising future directions that can further advance the research on WiFi sensing.

## VI. CONCLUSION

In this paper, we present a survey of WiFi sensing systems utilizing commodity devices. We first give a comprehensive overview of the WiFi sensing technical evolution. We then highlight the development of key techniques used in existing systems. Lastly, we conduct an in-depth analysis of the current challenges for WiFi sensing and possible future directions.

## ACKNOWLEDGEMENT

We thank the anonymous reviewers for their insightful feedback. This work was partially supported by the NSF Grants CNS 1910519, CNS 2131143, CNS 2120396, CNS1801630 and CCF2028876.

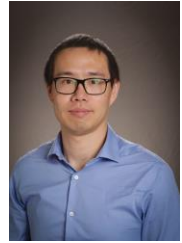

**Sheng Tan** received his Ph.D. degree in Computer Science from Florida State University in 2019. He is currently an Assistant Professor at the Department of Computer Science, Trinity University. His research interests include mobile computing, cyber security and human computer interaction.

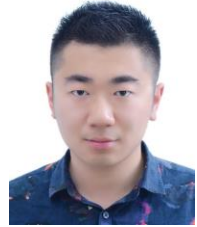

**Yili Ren** received the B.S. degree from Zhengzhou University in 2014, and the M.S. degree from Beihang University in 2017. He is currently pursuing a Ph.D. degree with the Department of Computer Science, Florida State University. His research interests include mobile computing, cyber security, and human-computer interaction.

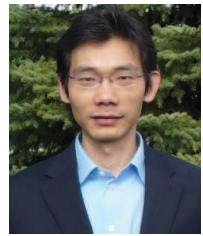

**Jie Yang** is an associate professor in the Department of Computer Science at Florida State University. He directs research in mobile computing and cybersecurity, is a pioneer in the area of WiFi sensing and mobile authentication. His work has been regularly featured in the media, including MIT Technology Review, NewScientist, Yahoo News, NPR, the New York Times, and The Wall Street Journal. He has published one book and three book chapters and 100+ research papers in prestigious journals and conferences. His recognitions include the FSU Developing Scholar Award, a Google Faculty Research Award, an FSU CS Faculty Research Award, the Stevens Francis T. Boesch Award, as well as Best Paper Awards at IEEE CNS 2014, IEEE CNS 2013, and ACM MobiCom 2011.

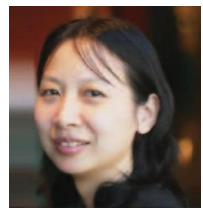

**Yingying (Jennifer) Chen** (S'94–M'01–SM'11) is a Professor of Electrical and Computer Engineering and Peter Cherasia Endowed Faculty Scholar at Rutgers University. She is the Associate Director of Wireless Information Network Laboratory (WINLAB). She also leads the Data Analysis and Information Security (DAISY) Lab. She is an IEEE Fellow and NAI Fellow. She is also named as an ACM Distinguished Scientist. Her research interests include mobile sensing and computing, cyber security and privacy, Internet of Things, and smart healthcare. She is a pioneer in the area of RF/WiFi sensing, location systems, and mobile security. She had extensive industry experiences at Nokia previously. She has published 3 books, 4 book chapters and 200+ journal articles and refereed conference papers. She is the recipient of seven Best Paper Awards in top ACM and IEEE conferences. Her research has been reported by numerous media outlets. She has been serving/served on the editorial boards of IEEE/ACM Transactions on Networking (IEEE/ACM ToN), IEEE Transactions on Mobile Computing (IEEE TMC), IEEE Transactions on Wireless Communications (IEEE TWireless), and ACM Transactions on Privacy and Security.